\documentclass{style}

\begin{document}

\title{a simultaneous 
solution scheme for coupled transonic accretion-wind systems}

\author{Tapas K. Das}

\address{IUCAA Post Bag4 Ganeshkhind Pune 411 007 INDIA
Email: tapas@iucaa.ernet.in
}




\maketitle

\abstracts{
We discuss a {\it non-self-similar analytical model} capable of explaining the 
formation of accretion-powered galactic and extra-galactic jets.
}
\vskip 0.5cm
\hrule
\noindent
{\bf Oral presentation, Session APT1. 
Published in the Proceedings of the IX th Marcel Grossmann Meeting.} 
\hrule

\section{Introduction}
Though widely observed to be emanating from a
variety of astrophysical sources, the underlying physical
mechanism behind the formation of galactic and extragalactic
jets is still enshrouded in a veil of mystery.  In
addition, it has not been possible to calculate accurately the
amount of barionic matter expelled in these events. Though progress in modelling
the exact mechanism of jet formation has been severely handicapped by 
inadequacy in understanding the proper microphysics responsible jet formation,
the basic ingradients of a theoretical model in attempt to explain the origin 
and formation of galactic/extragalactic jets seems to be well agreed upon
\cite{fer,mira,chak}. 
These are: firstly, a deep 
central potential well provided by an SMBH in case of jets from AGNs and QSOs
and stellar mass BHs in case of jets from galactic Microquasars, and secondly,
matter accreting onto these central compact objects to provide the energy for 
continuous throttling of the jets. Because of the fact that
BHs don't have their own
physical atmosphere from where matter could be ripped off as winds, outflows/jets
in these cases {\it have to be generated only from the accreting material}.
This invariable association of accretion processes with jet formation reinforce
the belief that for galactic and extra-galactic jet sources, inflow and outflow
must be studied within the same framework. Also to be noted that while self-similar
models are a valuable first step, they can  never be the full answer,
and indeed any model which works equally well at all radii is fairly
unsatisfactory to prove its viability. Thus the preferred model
for jet formation must be one which is able to
select the {\it specific} region of jet formation.\\
\section{Model Description}
Keeping the above mentioned facts at the back of our mind, we introduce
a {\it non
self-similar analytical model} 
which provides a simultaneous solution scheme for coupled inflow-outflow
systems to explain the hydrodynamic origin of accretion-powered
astrophysical outflows
\cite{d98,d99,d2000,d2000a,dc99}. 
By self-consistently combining the exact transonic accretion
and wind topologies, we compute the mass outflow rate $R_{\dot M}$
(the measure of the fraction of barionic accreting material being expelled
as outflows) {\it only} in terms of accretion parameters scaled in geometrical
unit and study the dependence of this rate on various physical
quantities governing the flow. We perform our calculation for
disk-outflow system as well as for accretion with negligible intrinsic
angular momentum. Unlike the self-similar solutions present in the
literature, our non-self similar model {\it for the
first time} points out the {\it exact}
location from where outflows are launched.\\
For the disc-outflow system, after explaining the fundamentals of the physical process by which the outflow
may generate from CENtrifugal Pressure Supported BOundary Layers 
(CENBOL) around
accreting compact objects, details of a mathematical scheme capable of
simultaneously solving the equations governing accretion and wind is provided
\cite{d98,d2000a,dc99}.
Connection between inflow-outflow topologies has been established along with the
self-consistent computation of the mass outflow rate $R_{\dot M}$. Also the
dependence of this rate on all possible accretion and shock parameters
has been thoroughly investigated.\\ 
Generation of outflows from spherical/ quasi-spherical
accretion onto non-rotating black holes has also been investigated
\cite{d99,d2000,d2000a}. Proposing
that a relativistic hadronic interaction supported steady and standing
spherical shock may be treated as the virtual hard surface from where the
outflow could be launched, mass outflow rate is computed from first
principle and its dependence
on various flow parameters have been studied by simultaneously solving the
inflow-outflow equatoins.\\
\section{Some Important Results and Scopes For Future Work}
For high compression ratio near the CENBOL as well as for outflows having 
large polytropic index, we show that runaway instability can take place
by rapid evacuation of accretion disc around the central accretor which 
indicates that our model is useful in explaining the quiescent states 
in X-ray novae systems. Also we show that a possible explanation of extremely
low luminosity and low radiative efficiency of our galactic centre could be 
due to the presence of profuse mass loss from near vicinity of
Sgr $A^{*}$ according to our model.\\
Recently it has been shown that significant neucleosynthesis can take place 
in accretion discs around black holes\cite{chakmukho}. We suggest that outflow 
generated according to our disc-outflow model would carry away
neucleosynthesis generated modified
compositions and could contaminate the atmosphere of the surrounding stars and
galaxies in general.

\end{document}